\documentclass[a4paper,fleqn,usenatbib]{mnras}

\usepackage{graphicx}	
\usepackage{amsmath}	
\usepackage{amssymb}	
\usepackage{subeqnarray}
\usepackage{hyperref}
\usepackage[para]{threeparttable} 
\usepackage[usenames, dvipsnames]{color}
\usepackage{longtable}
\usepackage[justification=centering]{caption}
\usepackage[T1]{fontenc}
\usepackage{ae,aecompl}
\usepackage{footnote}

\title[Luminosity Functions and Event Rate Densities of Long GRBs with Non-parametric Method]{A Comparative Study of Luminosity Functions and Event Rate Densities of Long GRBs with Non-parametric Method}

\author[Dong et al.]{
X. F. Dong$^{1}$,
X. J. Li$^{1}$,
Z. B. Zhang$^{1}$\thanks{E-mail: z-b-zhang@163.com},
and X. L. Zhang$^{1}$
\\
$^{1}$College of Physics and Engineering, Qufu Normal University, Qufu 273165, P. R. China\\
}
\date{Accepted XXX. Received YYY; in original form ZZZ}

\pubyear{2022}

\begin{document}
\label{firstpage}
\pagerange{\pageref{firstpage}--\pageref{lastpage}}
\maketitle

\begin{abstract}
In this work, we restudy the dependence of luminosity function and event rates for different gamma-ray burst samples on the criteria of sample selection and threshold effect. To compare with many previous studies, we have chosen two samples including 88 and 118 long bursts with known redshift and peak flux over 2.6 ph cm$^{-2}$ s$^{-1}$, from which 79 bursts are picked out to constitute our complete sample. It is found that the evolution of luminosity with redshift can be expressed by $L\propto(1+z)^k$ with a diverse $k$ relied more on the sample selection. Interestingly, the cumulative distributions of either non-evolving luminosities or redshifts are found to be also determined by the sample selection rather than the instrumental sensitivity. Nevertheless, the non-evolving luminosities of our samples are similarly distributed with a comparable break luminosity of $L_0\sim10^{51}$ erg s$^{-1}$. Importantly, we verify with a K-S test that three cases of event rates for the two burst samples evolve with redshift similarly except a small discrepancy due to sampling differences at low-redshift of $z<1$, in which all event rates show an excess of Gaussian profile instead of monotonous decline no matter whether the sample is complete. Most importantly, it is found that the burst rates violate the star formation rate at low redshift, while both of them are good in agreement with each other in the higher-redshift regions as many authors discovered previously. Therefore, we predict that two types of long bursts are favored in terms of their associations with both the star formation and the cosmic metallicity.
\end{abstract}


\section{Introduction }
\label{sec:Intro}
Gamma-ray bursts (GRBs) are the most energetic explosions found ever in the universe and produce huge amounts of energy in gamma-rays over a short time period ranging from a few milliseconds to thousands of seconds \citep{1993ApJ...413L.101K,2008A&A...484..293Z,2020ApJ...902...40Z}. They can even be detected at much higher redshifts than supernovae (SNe) that are generated from a stellar death. In theory, long GRBs (\textit{l}GRBs) with a duration $T_{\rm 90}>2$ s are believed to produce from core-collapsed massive stars (e.g., \citealt{1993AAS...182.5505W,1998ApJ...494L..45P, 2006ARA&A..44..507W}) which is evidently supported by observations of some GRBs associated with SNe, such as GRB 980425/SN 1998bw and GRB 030329/SN 2003dh (e.g., \citealt{2003Natur.423..847H,2003ApJ...591L..17S}).
The collapsar model implies that the GRB event rate should in principle trace the cosmic star formation rate (SFR; \citealt{1997ApJ...486L..71T,1998MNRAS.294L..13W,2000ApJ...536....1L,2001ApJ...548..522P,2004RvMP...76.1143P,2004IJMPA..19.2385Z,2007ChJAA...7....1Z}). It can be interestingly found that the low isotropic energy ($E_{\gamma,iso}$) SN/GRBs are relatively brighter in radio band compared to other long GRBs on a whole. According to \cite{1999ApJ...523..177W}, one can infer that the observed radio spectral peak luminosity ($L_{peak}$) of the SN/GRBs with smaller $E_{\gamma,iso}$ needs larger magnetic field ($B$) or larger number density ($n$) as $L_{peak} \sim n^{1/2}B^{1/2}$ in theory.

In the past two decades, many authors had focused on the study of relationship between the GRB event rate and the SFR in terms of different methods and samples, of which the direct fitting procedure with a specific function (e.g., \citealt{2007ApJ...662.1111L,2008ApJ...683L...5Y,2012ApJ...747...88N,2015MNRAS.448.3026W}) and the non-parametric method (e.g., \citealt{2012MNRAS.423.2627W,2015ApJ...806...44P,2015ApJS..218...13Y}) have been popularly adopted. However, the event rate estimates of long gamma-ray bursts based on distinct methods or samples especially at lower redshift are largely debated, parts of those results are contradictory with each other even though the same non-parametric method has been applied in literatures, which motivates us to revisit the dependence of luminosity function and event rates for different burst samples on sample selection and threshold effect systematically.
There are several algorithms to derive the luminosity function and event rate of GRBs for a specific kind of astronomical sources.
In fact, the observed GRB data are truncated in that the observational flux sensitivity of the satellite is limited. It is thus difficult to obtain a uniformly distributed GRB sample unless the selection effect is corrected.

Lynden-Bell's $c^{-}$ method \citep{1971MNRAS.155...95L,1992ApJ...399..345E} is one of the non-parametric and non-binning data processing techniques.
It was rediscovered by \cite{1985The Annals of Statistics..13..163} and \cite{1986The Annals of Statistics..14...1597}. This is not just an ordinary method but a unique nonparametric maximum likelihood estimator of randomly truncated univariate data, analogous to the famous Kaplan-Meier estimator for random censored data. It can readily combine samples with varied selection processes and is thus more powerful than the traditional fitting methods, which had let it be widely applied in the field of GRBs (e.g. \citealt{{2012MNRAS.423.2627W},{2015ApJS..218...13Y},{2015ApJ...806...44P},{2016A&A...587A..40P},{2017ApJ...850..161T},{2018ApJ...852....1Z},{2019MNRAS.488.5823L},{2021ApJ...920..135X}}). For example, \cite{{2015ApJS..218...13Y}} (hereafter Y15) adopted the non-parametric method for 127 GRBs and found that the event rate of GRBs decreases with the increase of redshift. While the SFR increases with redshift before $z\sim1$ and decreases with redshift after $z\sim1$, so that they claimed an excess of GRB event rate at low-redshift of $z<1$ (see also \citealt{2015ApJ...806...44P, 2018ApJ...852....1Z, 2019MNRAS.488.5823L}). On the contrary, \cite{2016A&A...587A..40P} (hereafter P16) utilized the same non-parametric method with a special sample selection to a sample of 81 \textit{Swift} \textit{l}GRBs and found no more excess of the GRB event rate than the SFR in the range of low-redshifts (see also \citealt{2012MNRAS.423.2627W}).
Recently, \cite{2019MNRAS.488.4607L} employed a maximum likelihood method to reexamine the luminosity function and event rate of 81 \textit{l}GRBs used in P16. They concluded that the GRB event rate may be consistent with the SFRs at $z<2$, but shows a discrepancy between them at $z>2$. Therefore, the relation of GRB event rate and the SFR especially at low-redshift end is still an open question. Unfortunately, such contradictions still can not be explained in theory reasonably. It is noticeable that the number of low-redshift GRBs in previous works is too limited, which may cause the estimate of event rate at lower redshift to be significantly biased.

To disclose the real evolution of GRB event rate with redshift, we will consider the same \textit{l}GRB samples but with different sample selection criteria and see how the GRB event rates evolve with redshifts diversely. Furthermore, we will expand the sample size of low-redshift GRBs in order to perform more reliable tests on the excessive component in statistics. In Section~\ref{sec:sample}, we describe how to build two \textit{l}GRB samples for three cases. In Section~\ref{sec:method}, we illustrate the non-parametric method and the data processing. Our results are presented in Section~\ref{sec:results}. Lastly, we end with conclusions in Section~\ref{sec:summary}.

\section{DATA}
\label{sec:sample}
\subsection{Sampling}

Since the launch of \textit{Swift} \citep{2004ApJ...611.1005G} and Fermi \citep{2009ApJ...702..791M} satellites, more and more GRBs with measured redshift are available recently, which is very helpful to investigate the evolution of luminosity with redshift completely. P16 pointed out that incompleteness of GRB samples will inevitably cause an excessive GRB event rate at low-redshifts because of the observational biases. To avoid the negative influences, they had chosen 81 \textit{Swift} \textit{l}GRBs with known redshift and higher peak photon flux than 2.6 ph cm$^{-2}$ s$^{-1}$ to re-constrain the GRB event rates at different redshifts. Strangely, they did not find the excessive components compared with the SFRs at lower redshift. In addition, \cite{2021MNRAS.504.4192B} recently argued that an underestimation of detection threshold will also lead to severely-incomplete \textit{l}GRB samples which eventually affects the inferred event rates.

Undoubtedly, the estimate of GRB event rate significantly depends on the sampling methods and/or the energy range of a detector in a certain sense.  Choosing two disparately selected samples together does not create a suitable sample for this kind of analysis. The issue has been recognised in the field and is part of the motivation for many more complete samples in the literatures such as TOUGH \citep{2012ApJ...756..187H}, BAT6 \citep{2012ApJ...749...68S} amd SHOALS \citep{2016ApJ...817....7P}. To check whether the excess of GRB event rate at low-redshift is biased by the effects of sample selection and threshold, it is necessary to give a comparative study for a united sample of \textit{Swift} \textit{l}GRBs complied with distinct sensitivities, on condition that the GRB samples should be uniformly built according to both flux and redshift in a full redshift coverage. For this purpose, we also adopt the lower flux limit of 2.6 ph cm$^{-2}$ s$^{-1}$ as our basic sampling criterion. Firstly, we pick 88 bright bursts out of 127 \textit{l}GRBs with both redshift and good spectral parameter from Y15 to comprise our sample I. Secondly, we add 30 low-redshift ($0<z<1$) \textit{l}GRBs published in \citep{2018PASP..130e4202Z} to build our sample II (N=118) in order to compensate the number deficiency of low-redshift \textit{l}GRBs of previous works \citep[e.g.][]{2012ApJ...749...68S,2016A&A...587A..40P}, where brighter LGRBs with larger luminosities or redshifts were easily chosen in that well-constraints on Ep of the GRB spectra naturally lead to  higher factions of high-luminosity LGRBs and some less-luminous LGRBs with smaller redshift had been neglected. Besides, the redshift coverage should be wide enough and the sufficient low-redshift bursts are essential to warrant more credible estimates of event rate as a whole for a complete sample. We need to point out that all bursts but GRB 090328 in our sample were detected by \textit{Swift} satellite initially. In addition, 31 and 62 out of 118 \textit{l}GRBs are also successfully observed by Ferm/GBM and Konus-wind individually.

\begin{figure}
	\centering
	\includegraphics[width=0.45\textwidth]{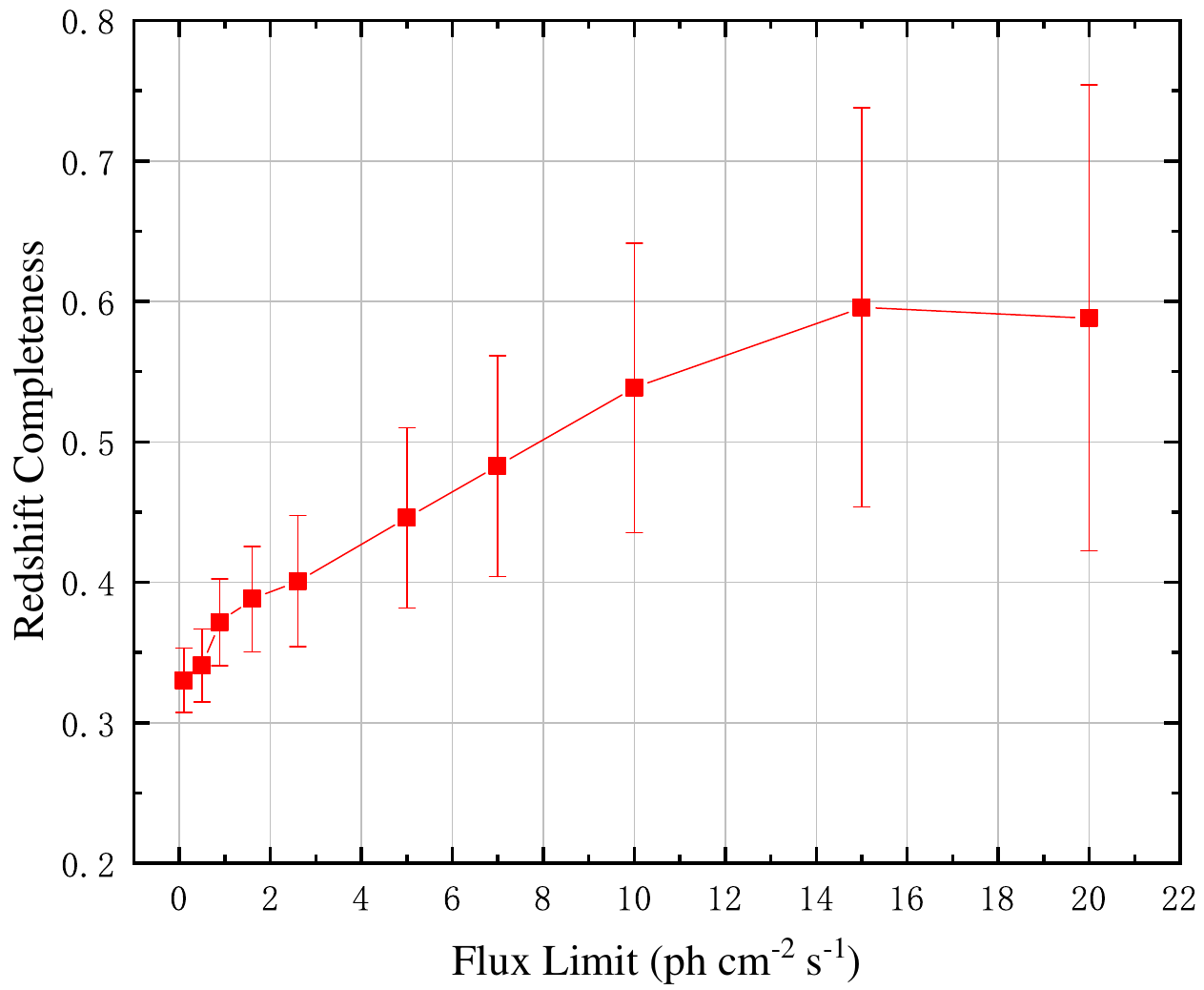}
	\caption{\protect{Redshift completeness degree versus peak photon flux cut for 854 \textit{Swift} \textit{l}GRBs, of which 280 bursts had been measured with known redshift. A Poisson error has been given to each data point by error propagation.}}
	\label{fig0}
\end{figure}

Considering the complete conditions of sample selection defined by \cite{2006A&A...447..897J} and \cite{2012ApJ...749...68S} carefully, we pick out 54 and 79 \textit{l}GRBs that meet the completeness standards and estimate the completeness degrees to be about 54/88$\sim$61\% and 79/118$\sim$67\% for our samples I and II, respectively. Regarding the 30 low-redshift LGRBs from \cite{2018PASP..130e4202Z}, the completeness goes up to 25/30$\sim$83\%. In addition, we also collect 854 \textit{l}GRBs with peak fluxes recorded on the official \textit{Swift} website from Dec 2004 to Sep 2018, of which 280 \textit{l}GRBs with known redshift, and examine how the redshift completeness depends on different flux cut in Fig. \ref{fig0}. It can be seen that the redshift completeness will go up with increase of flux cuts on condition that the remaining number of bursts above a certain flux cut is large enough. At most, the completeness level can reach no more than 0.6 if the above complete conditions are dismissed. Meanwhile, the completeness uncertainty becomes more and more large as the redshift increases. Some parameters of the two samples are presented in Table 1 as GRB name (Column 1), redshift $z$ (Column 2), $T_{90}$ (Column 3), peak flux $P$ (Column 4) in a certain energy range (Column 5), bolometric luminosity $L$ (Column 6) and references (Column 7). We caution that all \textit{l}GRBs in our sample have been strictly chosen by taking into account the P16 sample selection criteria \citep[see also][]{2006A&A...447..897J,2012ApJ...749...68S} and marked with a star symbol after GRB name. As a result, 79 \textit{l}GRBs with both constraints of the complete conditions and good spectra are selected to built our complete sample III. Note that the fraction of low-reshift bursts with $z<1$ is as high as about 42 percent in our whole sample, which guarantees that the GRB event rates at lower redshift can be really and adequately reproduced. Moreover, our sample expansion at low redshift not only warrants an adequate redshift coverage but also ensures that the sample is uniformly built regarding fluxes in a Euclidean space.

\subsection{Luminosity Limit}

The peak luminoity of the GRBs is calculated by $L=4{\pi}d_{L}^{2}(z)FK$, where $F$ is the observed peak flux within a certain energy range and $K$ denotes the factor of $K$-correction factor \citep{2018PASP..130e4202Z}. The luminosity distance $d_{L}(z)$ at a redshift z \citep{1999astro.ph..5116H} is written as
\begin{equation}
d_{L}(z)=\frac{c}{H_0}(1+z)\int_0^z
\frac{dz}{\sqrt{1-\Omega_{m}+\Omega_{m}(1+z)^3}}.
\end{equation}
Throughout the paper, a flat $\Lambda$CDM universe with
${\Omega}_{m}=0.27$, ${\Omega}_{\Lambda}=0.73$ and $H_{0}=70~{\rm km~s^{-1}Mpc^{-1}}$ has been assumed.

It is known that the flux threshold of \textit{Swift}/BAT is not straightforward to determine precisely since many triggers on the spacecraft make the sensitivity of detectors very complex for parameterization well \citep{2006ApJ...644..378B}. Following \cite{2004ApJ...611.1005G}, \cite{2008MNRAS.388.1487L} and \cite{2015ApJS..218...13Y}, we assume the instrumental sensitivity to be about $F_{\rm {lim,1}}=2.0\times10^{-8}~{\rm erg~cm^{-2}s^{-1}}$ for our calculation subsequently. The luminosity limit at a certain redshift $z$ can be given as $L_{\rm lim} =4{\pi}d_{\rm L}^{2}(z)F_{\rm lim}$ in that the K-correction parameter is narrowly distributed around $1$ in view of precious investigations (see e.g. \citealt{2003ApJ...594..674B,2018PASP..130e4202Z}).
On the other hand, \cite{2021MNRAS.504.4192B} pointed out that the detection threshold effect will be underestimated in a sense and should be given a conservative estimation of
$F_{\rm {lim,2}}=1.0\times10^{-7}~{\rm erg~cm^{-2}s^{-1}}$. The difference between $F_{\rm {lim,1}}$ and $F_{\rm {lim,2}}$ could play an un-negligible role on modeling the observed luminosity-redshift relation when a significant fraction of \textit{l}GRBs reside below the $L_{lim}$ for a GRB sample. Therefore, we will pay more attentions to the influence of not only the sample selection but also the threshold effect on the GRB luminosity evolving with redshift for the above two refined \textit{l}GRBs samples.

\section{METHOD}

\label{sec:method}
The Lynden-Bell's $c^{-}$ method adopted here requires the luminosity $L$ is independent of the redshift $z$ in advance \citep{{1971MNRAS.155...95L},{1992ApJ...399..345E}}, so that the luminosity function and the GRB event rate can be accurately determined. Therefore, we need to reduce the redshift evolution effect of $L$ on $z$ with a nonparametric $\tau$ test method firstly.

\subsection{The method of $\tau$ statistics}

\begin{figure}
	\centering
	\includegraphics[width=0.5\textwidth]{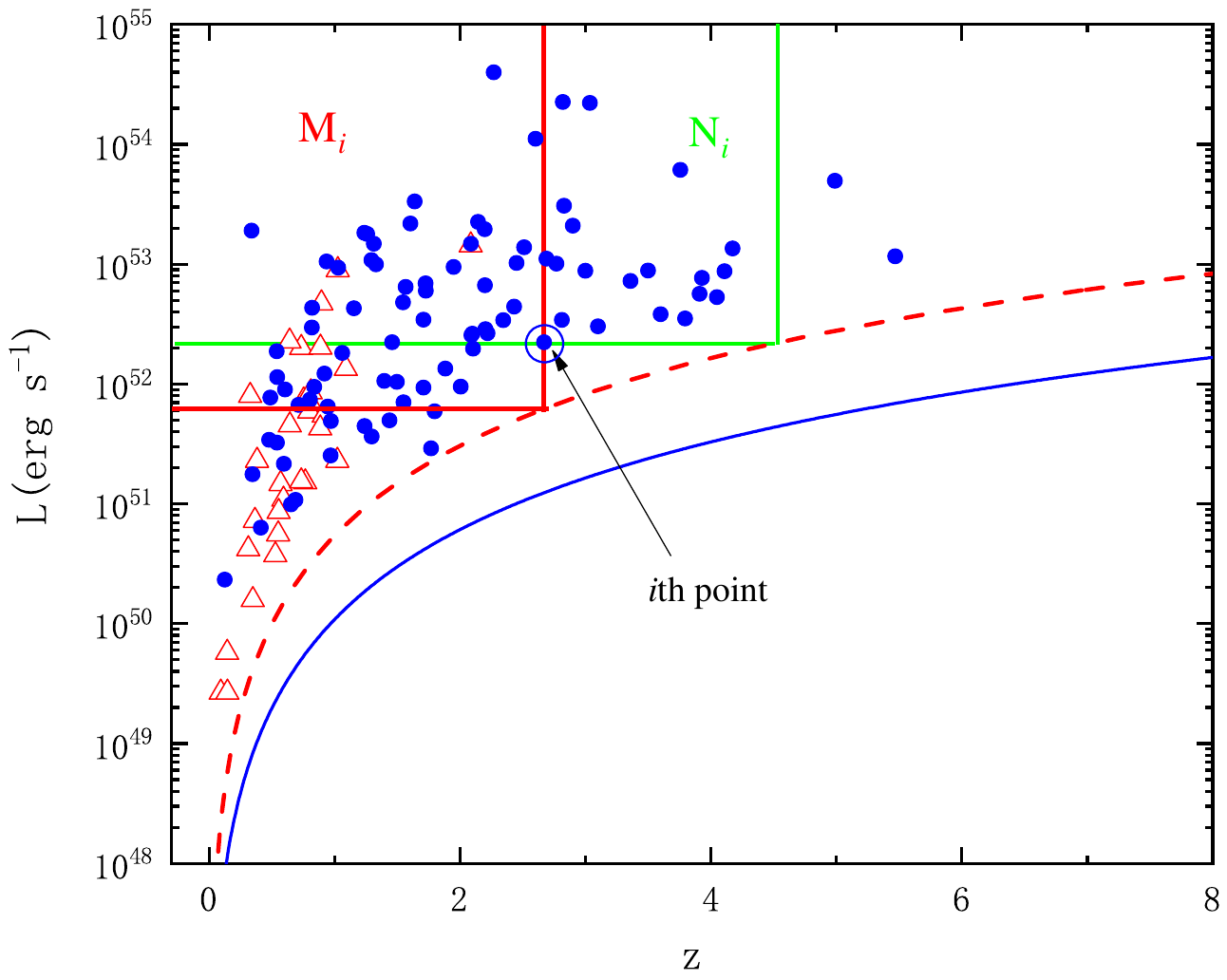}
	\caption{\protect{The filled circles represent the 88 bright \textit{l}GRBs. The empty triangles represent 30 low-redshift \textit{l}GRBs taken from \citet{2018PASP..130e4202Z}. The dashed and solid lines represent the luminosity limits estimated for the flux sensitivities of $F_{lim,1}$ and $F_{lim,2}$, respectively.}}
	\label{fig1}
\end{figure}

If $L$ and $z$ are independent of each other then one can write the joint distribution as $\Psi(L,z)=\psi(L)\phi(z)$, in which $\psi(L)$ is a luminosity function of the GRBs and $\phi(z)$ represents the cumulative redshift distribution \citep{1992ApJ...399..345E}. Observationally, the  luminosity is positively correlated with the redshift for our samples as shown in Figure~\ref{fig1}. As usual, $\Psi(L,z)$ can be decomposed into the form of $\Psi(L,z)=\psi(L/g(z))\phi(z)$, where $g(z)$ describes the evolutive relationship between $L$ and $z$. And if letting $L_{0}=L/g(z)$, we then get $\Psi(L_0,z)=\psi(L_0)\phi(z)$, of which the redshift $z$ and the modified luminosity $L_0$ are independent and already satisfy the requirement of the non-parametric $\tau$ test.

\begin{figure*}
\begin{center}
\includegraphics[width=1.0\linewidth, angle=0,scale=1]{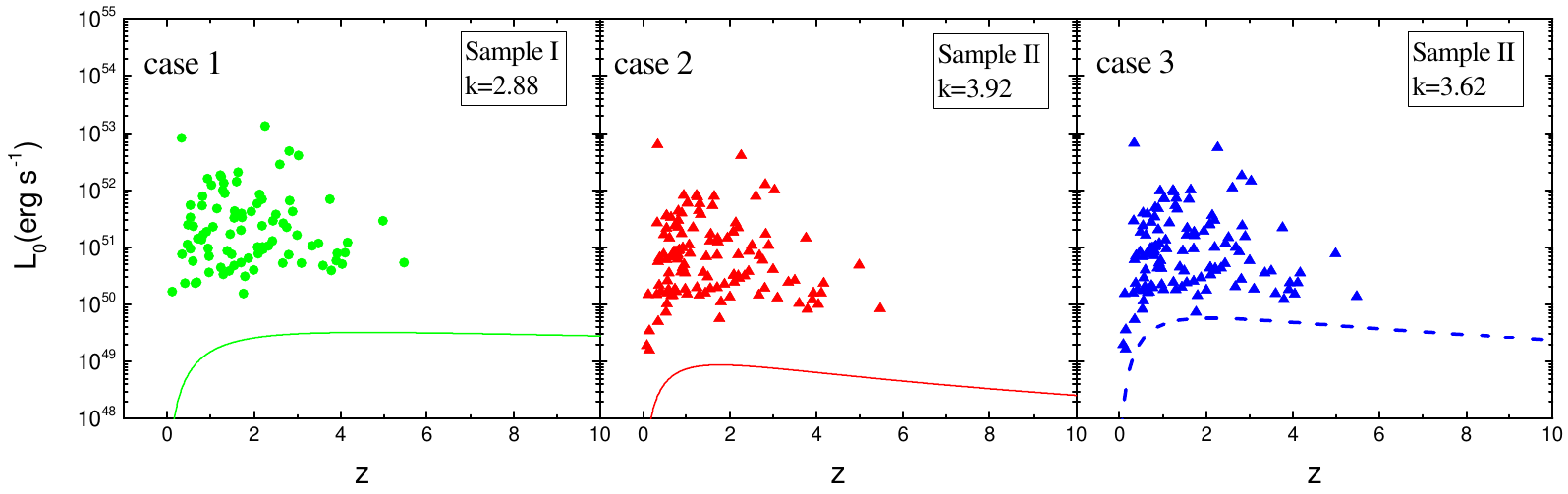}
\vskip-0.1in
\caption{The relationships between $z$ and $L_0$ for samples I (in case 1) and II ( in cases 2 and 3), where the solid and the dashed lines represent the lower luminosity limits decided by two sensitivities of $F_{\rm {lim,1}}$ and $F_{\rm {lim,2}}$, respectively.
}
\label{fig2}
\end{center}
\end{figure*}

We continuously adopt the power-law form of $g(z)=(1+z)^k$ that has been used in many literatures (e.g. \citealt{2002ApJ...574..554L,2004ApJ...609..935Y,2015ApJ...800...31D,2015ApJ...806...44P,2015ApJS..218...13Y,2018ApJ...852....1Z}). First, suppose that we get a definite value of $k$ and after removal, each data point changes from $(z_i,L_i)$ to $(z_i , L_{0,i})$. For the $i$th data in the $(z_i,L_{0,i})$ data set, we can define $J_i$ as
\begin{equation}\label{Ji}
J_i = \{j|L_{0,j} \geq L_{0,i} , z_j\leq z_{i}^{max}\},
\end{equation}
where $L_{0,i}$ is the $i$th GBR luminosity without redshift evolution and $z_{i}^{\rm max}$ is the maximum redshift at which a GRB with luminosity $L_{0,i}$ can be observed. The number of GRBs contained in this region is $n_i$. The number of GRBs with redshift $z$ less than or equal to $z_i$ in this region is defined as $R_i$. The $\tau$ test statistic is defined to be
\begin{equation}
\tau \equiv \frac{\sum_{i}(R_i - E_i)}{\sqrt{\sum_{i}{V_i}}},
\end{equation}
where $E_i = \frac{1+n_i}{2}$, $V_i = \frac{(n_i- 1)^2}{12}$ are the expected mean and the variance of $R_i$, respectively. As known from the $\tau$ test statistic, if $R_i$ is exactly uniformly distributed between 1 and $n_i$ then the sample number of $R_i \leq E_i$ and $R_i \geq E_i$ should be nearly equal and the value of $\tau$ will be nearly 0, then $L_0$ and $z$ become independent of each other after removing the evolution with $g(z)=(1+z)^k$. Based on this, we have to adjust the value of $k$ until the $\tau$ is equal to 0 from which we can get the expected value of $k$ in $g(z)$.

Subsequently, we constrain the $k$ values for the distinct samples I and II at $1\sigma$ confidence level. When the $F_{\rm lim,1}$ is used, the best $k$ values are $2.88_{-0.45}^{+0.41}$ and $3.92_{-0.37}^{+0.38}$ for samples I and II. Surprisingly, the $k$ index of sample II will become $3.62_{-0.34}^{+0.42}$ when the tighter flux cut of $F_{\rm lim,2}$ is applied. It demonstrates that the deduced $k$ values depend more on the sample selection but less on the instrumental effect, which is somewhat different from what mentioned by \cite{2021MNRAS.504.4192B}.
The reason is that all bursts in our samples are located above the lower limits of luminosities to ensure the completeness of our samples. It is noticeable that the index $k=2.88_{-0.45}^{+0.41}$ of our sample I is consistent with most previous results (see e.g., \citealt{2012MNRAS.423.2627W}, \citealt{2016A&A...587A..40P} and \citealt{2015ApJS..218...13Y}), but larger than those values of 1.7 and 1.2 given by \cite{2017ApJ...850..161T} and \cite{2021ApJ...908...83T}, respectively. In contrast, the $k$ indexes of our sample II are much closer to $k={3.5\pm0.5}$ provided by \cite{2019MNRAS.488.5823L}. Consequently, one can conclude that the diverse $k$ values reported in many literatures are changing on a relatively large scale from sample to sample and confirm again that the sample selection effect does influence the determination of non-evolving luminosites and GRB event rates in evidence.

For simplification, we define cases 1, 2 and 3 to represent sample I with $F_{\rm {lim,1}}$, sample II with $F_{\rm {lim,1}}$ and sample II with $F_{\rm {lim,2}}$, respectively. Figure~\ref{fig2} shows the relationships between $z$ and $L_0$ after the redshift evolution of $g(z)$ was removed by $L_0=L/(1+z)^{k}$ for the above three cases. It can be clearly seen that the non-evolving luminosity $L_0$ of two above samples is already independently with redshift. In the following, we will utilize the data of $L_0$ and $z$ to derive the model-independent luminosity functions and event rates of two samples of \textit{l}GRBs.

\subsection{Luminosity Function and Event Rate of GRBs}

The Lynden-Bell's $c^{-}$ method is an effective way to determine the redshift distribution and luminosity function of astronomical objects using the truncated samples.
Let
\begin{equation}
N_i = n_i - 1
\end{equation}
represent the number of GRBs contained in $J_i$ which can be understood as the minus one count (taking the $i$th point out) called as the Lynden-Bell's $c^{-}$ method \citep{1971MNRAS.155...95L}.
And we then set
\begin{equation}\label{JI}
J^{\prime}_i = \{j|L_{0,j} \geq L_{0,i}^{lim} , z_j < z_{i}\},
\end{equation}
and let $M_i$ to be the number of \textit{l}GRBs contained in $J^{\prime}_i$.
The cumulative luminosity function can be derived from the following formula by the nonparametric method \citep{1971MNRAS.155...95L,1992ApJ...399..345E}:
\begin{equation}\label{LuminosityFunction}
\psi(L_{0,i}) = \prod\limits_{j<i}(1+\frac{1}{N_j}),
\end{equation}
where $j<i$ means that GRB has luminosity $L_{0,j}$ larger than
$L_{0,i}$.
The cumulative redshift distribution $\phi(z)$ can be
obtained from
\begin{equation}\label{redshiftFunction}
\phi(z_i) = \prod\limits_{j<i}(1+\frac{1}{M_j}),
\end{equation}
where $j<i$ means that GRB has redshift $z_j$ less than $z_i$.
The event rate of GRBs can be written as
\begin{equation}\label{formationrate}
\rho(z) = \frac{d\phi(z)}{dz}(1+z)(\frac{dV(z)}{dz})^{-1},
\end{equation}
where $(1+z)$ results from the cosmological time dilation and
$dV(z)/dz$ is the differential comoving volume which can be expressed as \citep{2019JHEAp..24....1K}
\begin{equation}\label{comovingvolume}
\frac{dV(z)}{dz}=\frac{c}{H_0}\frac{4\pi{d^2_{l}(z)}}{(1+z)^2}\frac{1}{\sqrt{1-\Omega_{\rm m}+\Omega_{\rm m}(1+z)^3}},
\end{equation}
where the comoving volume at a redshift of $z$ is $V=4\pi{D_M^3}/3$ with the comoving distance of $D_M=d_l/(1+z)$ (\citealt{1999astro.ph..5116H}).

\section{Results}
\label{sec:results}
In this section, we give our results of luminosity functions and event rate of \textit{l}GRBs constrained by the non-parametric method. Simultaneously, we compare the evolutionary history of distinct \textit{l}GRB event rate with that of star formations.

\subsection{Luminosity functions of different \textit{l}GRB samples}

\begin{figure}
\includegraphics[width=0.9\columnwidth]{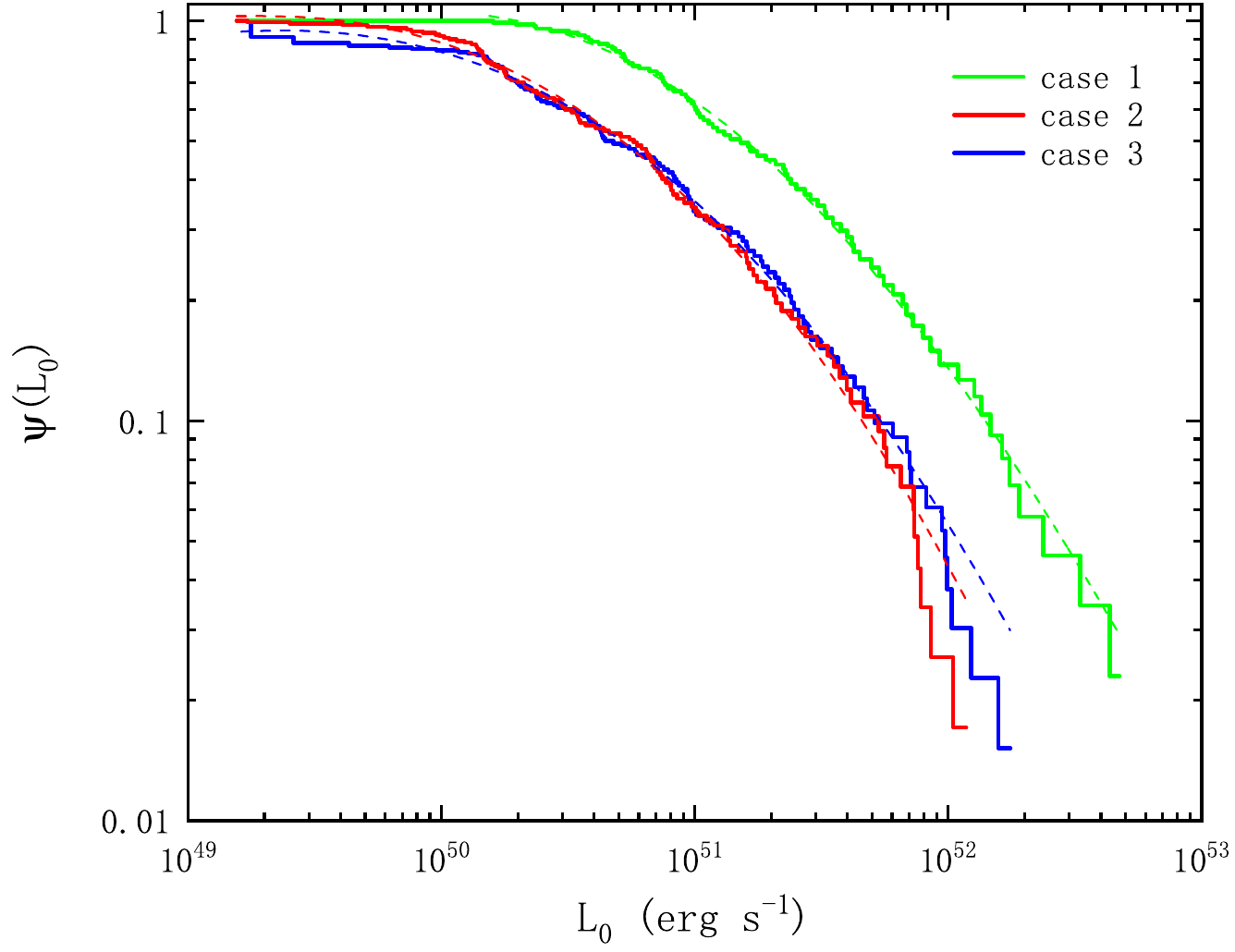}
\centering
\caption{Cumulative luminosity functions of two samples of \textit{l}GRBs. The green line represents the case 1, the red line represents the case 2 and the blue line represents the case 3. All data are corrected to the first point. The corresponding dashed lines represent the best fits in each case.
}
\label{fig3}
\end{figure}

Using the Lynden-Bell $c-$ method in Eq.(\ref{LuminosityFunction}), we now obtain the cumulative luminosity functions of two different \textit{l}GRB samples with distinct sensitivities. Figure~\ref{fig3} depicts that the normalized cumulative luminosity functions decrease gradually with the increase of non-envolving luminosity, which is similar to some previous studies \citep{2015ApJS..218...13Y,2016A&A...587A..40P,2017ApJ...850..161T,2019MNRAS.488.4607L}. It is noticeable that the luminosity function in case 1 is significantly different from those in both cases 2 and 3, while the cases 2 and 3 are largely consistent with each other. This indicates that the derived luminosity function is indeed sensitive to the sample selection other than the sensitivity dramatically provided that no GRBs appear below $L_{lim}$ in Figure~\ref{fig1}.

We now fit the cumulative luminosity distribution in each case with a smoothly broken power-law function defined by \begin{equation}\label{fit}
\psi(L_0) =\psi_{\ast}[(\frac{L_0}{L_b})^{\alpha\omega}+(\frac{L_0}{L_b})^{\beta\omega}]^{-\frac{1}{\omega}},
\end{equation}
in which $L_b$ is the break luminosity and $\psi_{\ast}$ is a normalization factor, $\alpha$ and $\beta$ are two power-law indexes characterizing the decay of luminosity function before and after the $L_b$, and $\omega$ is a smoothness parameter assigned to be $0.18$ empirically in this study. As a result, we get the break luminosity $L_b$ and other fitted parameters to be $L_{b,1}=(8.5\pm2.9)\times10^{50}~{\rm erg\ s^{-1}}$ for $\alpha=-0.63\pm0.02$, $\beta=1.46\pm0.02$ and $\psi_{\ast}=30.52\pm0.12$ in case 1, $L_{b,2}=(2.5\pm1.6)\times10^{51}~{\rm erg\ s^{-1}}$ for $\alpha=1.97\pm0.04$, $\beta=-0.27\pm0.01$ and $\psi_{\ast}=7.86\pm0.15$ in case 2 and $L_{b,3}=(1.8\pm1.3)\times10^{51}~{\rm erg\ s^{-1}}$ for $\alpha=1.73\pm0.02$, $\beta=-0.34\pm0.01$ and $\psi_{\ast}=11.01\pm0.13$ in case 3, respectively. Interestingly, the three break luminosities are very close although the luminosity distribution of case 1 is obviously different from those of both cases 2 and 3. It is worthy to emphasize that the break luminosities are not affected by the threshold effect and the sample selection remarkably.

\subsection{Event rate densities of \textit{l}GRBs}
\subsubsection{Comparison between different samples}
\begin{figure}
\includegraphics[width=0.9\columnwidth]{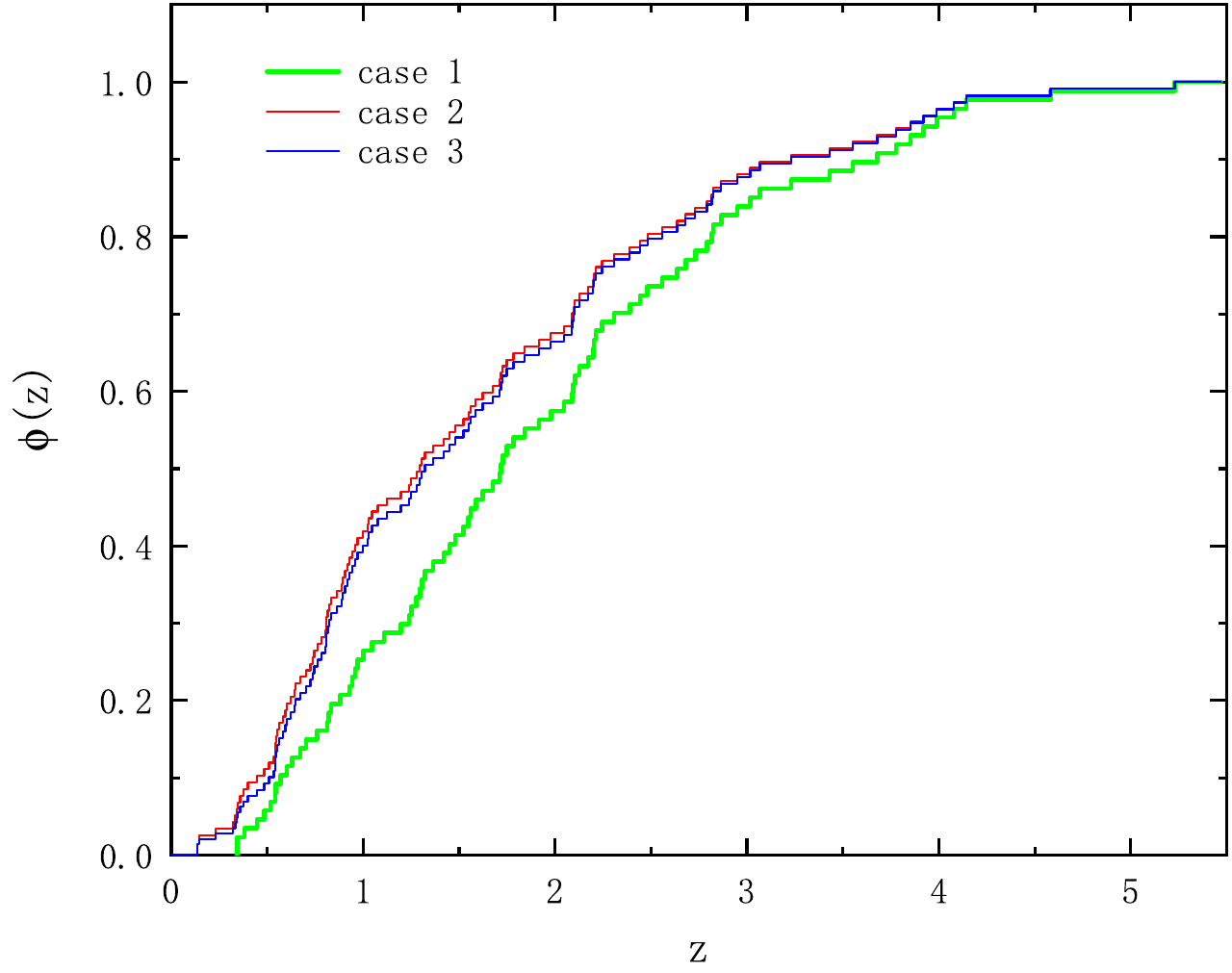}
\centering
\caption{Cumulative redshift distributions of two samples of \textit{l}GRB. The green line represents the case 1, the red line represents the case 2, and the blue line represents the case 3. All data are normalized to the maximum redshift point.
}
\label{fig4}
\end{figure}

Figure~\ref{fig4} displays the normalized cumulative redshift distributions of \textit{l}GRBs for samples I and II from Eq. (\ref{redshiftFunction}). It can be seen that $\phi(z)$ increases gradually with redshift, which is consistent with some previous studies \citep[e.g.,][]{2012MNRAS.423.2627W,2015ApJ...806...44P,2015ApJS..218...13Y}.
Especially, the cumulative $\phi(z)$ functions of cases 2 and 3 are found to evolve with redshift in a similar way. In contrast, the cumulative $\phi(z)$ function in case 1 behaves smoother at lower redshift and steeper at higher redshift. Figure~\ref{fig4} also demonstrates that the cumulative redshift distributions are almost unaffected by the sensitivity of detectors while evidently biased by the effect of sample selection which is perfectly consistent with what illustrated in Figure~\ref{fig3}. Noticeably, a slope transformation of the redshift distribution function in Figure~\ref{fig4} will lead to distinct evolutions of GRB event rate with the cosmological redshift according to Eq. (\ref{formationrate}).

Three ladder lines in Figure~\ref{fig5} correspond to the three cases of GRB event rates $\rho(z)$ evolving with redshift. It is worthy to point out that all GRB event rates derived from our samples show an excess of Gaussian profile compared with the SFRs at lower redshift of $z<$1, which vastly differs from either the monotonous decline proposed by many authors \citep[e.g.,][]{2015ApJS..218...13Y,2015ApJ...806...44P,2019MNRAS.488.5823L,2021ApJ...908...83T} or the monotonous rise presented by \citep{2004ApJ...609..935Y,2012MNRAS.423.2627W,2016A&A...587A..40P,2019MNRAS.488.4607L}. It is because the number of low-redshift bursts in previous papers is too limited to manifest the real evolutional profile of GRB event rates. If a significant fraction of low-redshift GRBs with less brightness are included, the GRB event rates will also go up at lower-redshift region. However, the sample was already seriously biased in this situation, which may cause an untrustworthy estimate of GRB event rate.

Again, one can find that the GRB event rates are not affected by the threshold significantly but depend more on the sample selection instead.
A Kolmogorov-Smirnov (K-S) test to any two cases of GRB event rates returns $D_{12}\simeq0.31$ $ (p=0.53)$, $D_{23}\simeq0.23 $ $(p=0.86)$ and $D_{13}\simeq0.27 $ $(p=0.70)$, respectively. If adopting the critical value $D_{\alpha}\sim0.64$ at a significance level of $\alpha=0.01$ (far less than $p$-values), we can therefore conclude that the three cases of GRB event rates are surprisingly taken from the same distribution no matter how to comprise the complete samples. Since these GRB events are transients, their relevant density is in space-time, not just space. One need to correct the inferred event rate via dividing the detected number of GRBs by actual observation time and sky coverage, instead of the total time and space.

\subsubsection{GRB event rate versus star formation rate}
\begin{figure*}
\begin{center}
\includegraphics[angle=0,scale=0.5]{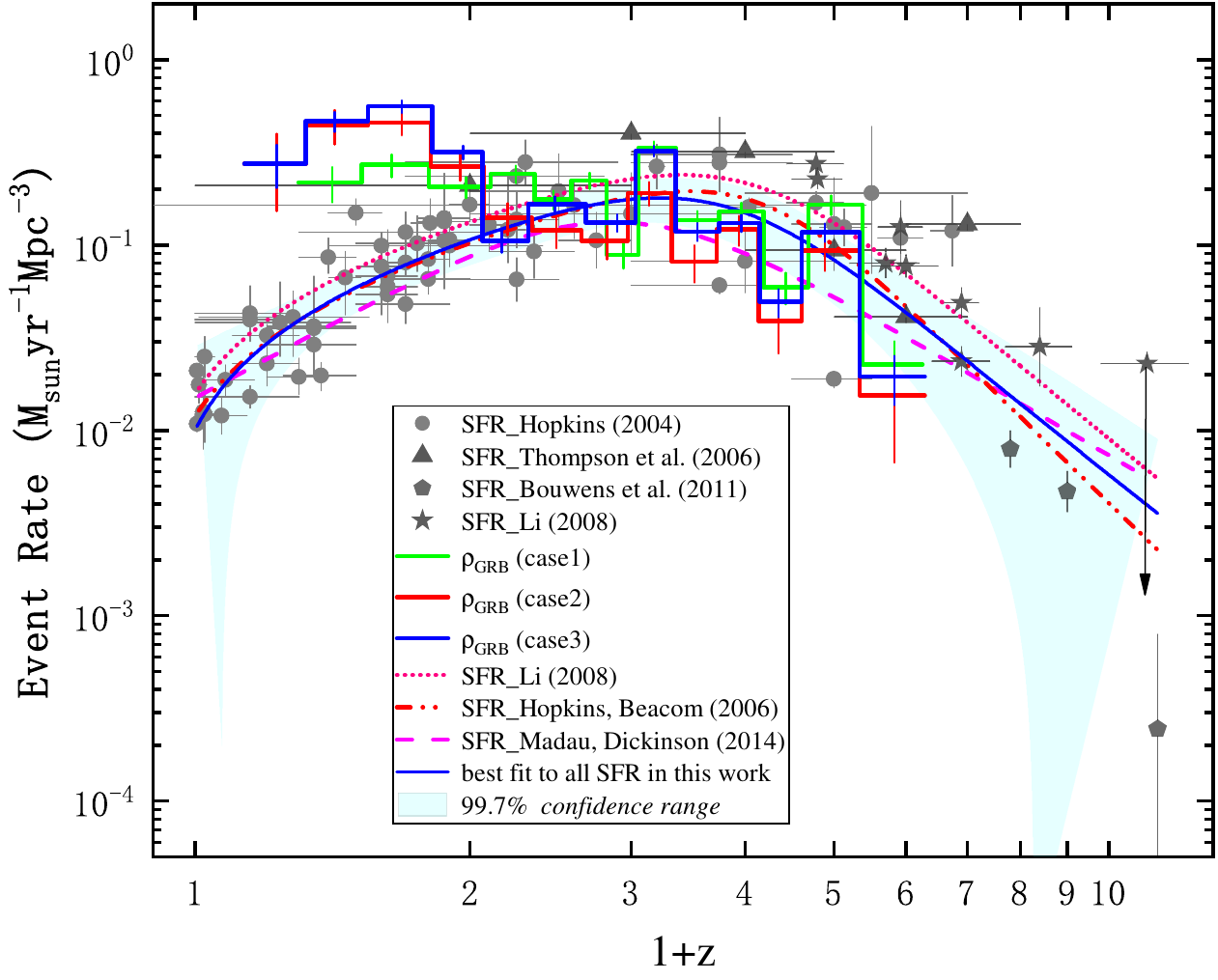}
\vskip-0.1in
\caption{
Comparison of GRB event rates in cases 1, 2 and 3 with SFR. The green, red and blue ladder lines represent the evolutions of GRB event rate changing with redshift for the case 1, 2 and 3, respectively. Gray dots, triangles, pentagons and stars represent the observed SFRs recorded in \citet{2004ApJ...615..209H}, \citet{2006ApJ...647..787T}, \citet{2011Natur.469..504B} and \citet{2008MNRAS.388.1487L}, respectively. The pink dashed, red double-dot dashed, and magenta dashed lines correspond to the theoretical lines of the SFR evolving with the redshift in \citet{2008MNRAS.388.1487L}, \citet{2006ApJ...651..142H}, \citet{2014ARA&A..52..415M}, respectively, while the blue solid line represents our best fit to all the SFR data with a 99.7$\%$ confidence level.
}
\label{fig5}
\end{center}
\end{figure*}

\cite{2001ApJ...548..522P} pointed out that the event rate of GRB traces the global star formation history of the universe. It was usually assumed that the GRB event rates is proportional to the SFRs in literatures \citep[e.g.,][]{2001ApJ...548..522P,2019MNRAS.488.5823L,2019MNRAS.488.4607L,2021A&A...649A.166P}, which enables us to compare our newly-built GRB event rates with the SFRs constrained by the largest sample of stars observed within a wider redshift range ever in Figure~\ref{fig5}, where we find that all three GRB event rates trace the SFRs well at higher redshift of $z>1$ and exhibit an obvious excess with a Gaussian-like shape at $z<1$. This strongly indicates that the \textit{l}GRBs should be classified into two groups depending on whether they match the SFRs or not. Hence we can optimistically propose that the low-redshift \textit{l}GRBs with higher event rates are not associated with the SFRs and should stand for a separate subclass. Note that the low-redshift excess of case 1 is not significant because the fraction of low-redshift \textit{l}GRBs in case 1 is relatively lower.

\begin{figure*}
\begin{center}
\includegraphics[angle=0,scale=0.5]{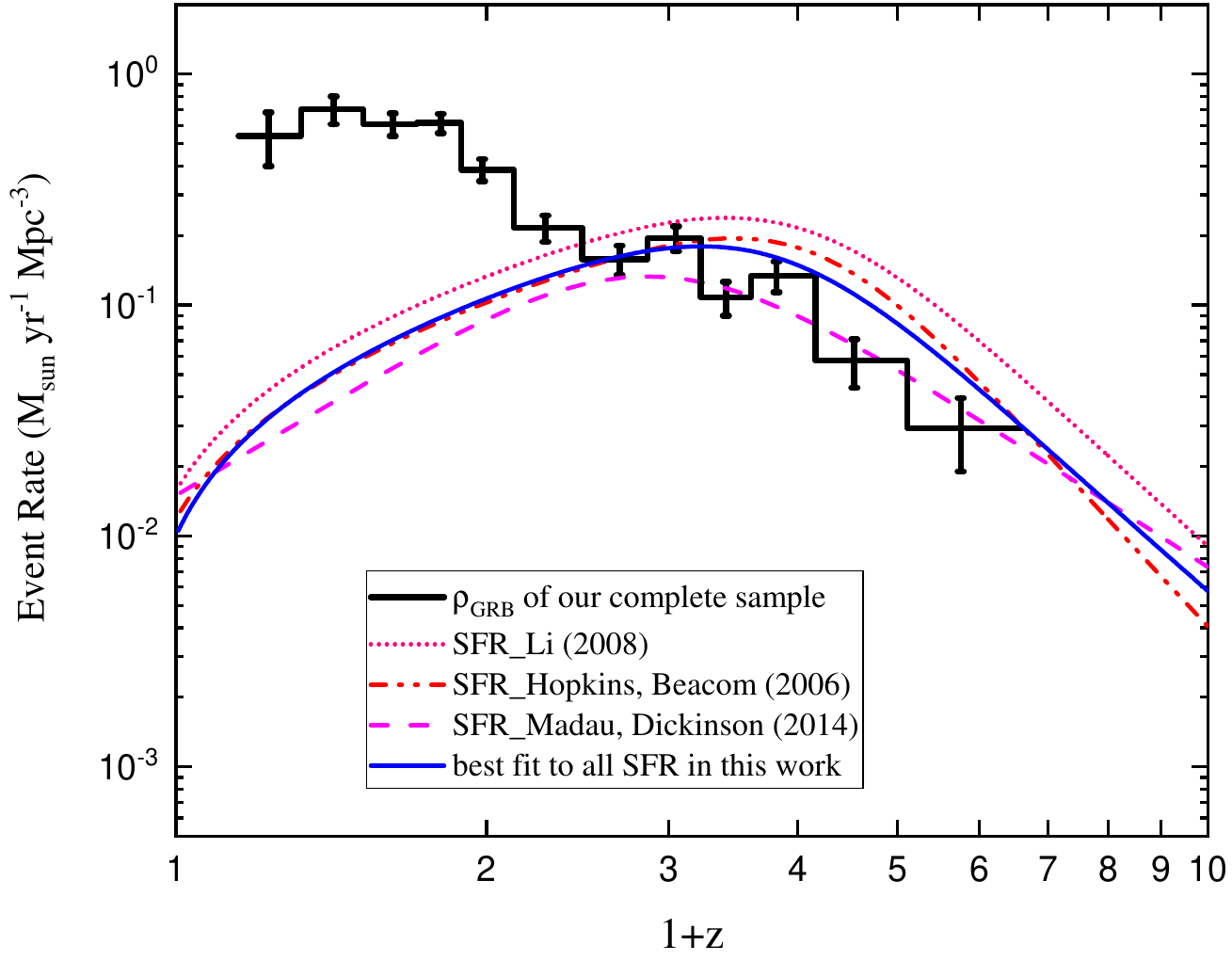}
\vskip-0.1in
\caption{Comparison of GRB event rate of sample III with different SFR fitting lines. The black solid ladder line stands for GRB event rate of our complete sample evolving with redshift. All curved symbols are same as in Figure \ref{fig5}.}
\label{fig6}
\end{center}
\end{figure*}

Now, we calculate the GRB event rate of sample III including 79 \textit{l}GRBs matching the complete constraints like P16 and contrast with different SFR models in Figure \ref{fig6}, from which we surprisingly find that the discrepancy between GRB event rate and SFRs becomes more evident at low redshift. On the contrary, the concordant tendency proposed by P16 can not be recovered at the low-redshift region anyhow. Most probably, P16 had   omitted the term of $(dV/dz)^{-1}$ from Eq. (\ref{formationrate}) in their calculation. Previously, Y15 also noticed the mistake appeared in \cite{2004ApJ...609..935Y} and \cite{2012MNRAS.423.2627W}, which led to a problematic rising trend of GRB event rate matching SFR at low redshift. It is worth noting that the same non-parametric method has been adopted in these researches, but Y15, \cite{2004ApJ...609..935Y} and \cite{2012MNRAS.423.2627W} did not apply to a complete \textit{l}GRB sample at all.

Regarding the SFRs themselves, we use the common form of $\dot{\rho_{\ast}}=(a+bz)h/[1+(z/c)^d]$ with $h=0.7$ \citep{2006ApJ...651..142H} to get the best fitting parameter set ($a=0.014\pm0.009$, $b=0.140\pm0.018$, $c=2.98\pm0.22$ and $d=4.55\pm0.54$). The best fit to the updated SFR data in the work has been highlighted with solid line in Figure~\ref{fig5} and \ref{fig6}, from which one can find our results are coincident with those previous ones \citep[for example][]{2004ApJ...615..209H,2006ApJ...647..787T,2008MNRAS.388.1487L}. It was demonstrated by \cite{2008MNRAS.388.1487L} that \textit{l}GRBs at high redshift always traced the SFR and the comic metallicity evolution on basis of theories and observations \citep[see e.g.,][]{2006MNRAS.372.1034D,2008ApJ...673L.119K,2021A&A...649A.166P,2021arXiv211200643F}. In any case, however, we find that the SFRs are always lower than GRB event rates regardless of the completeness of GRB samples in the low-redshift range. In fact, \cite{2016SSRv..202..181C} showed that the GRB event rates per unit star-formation at z>3 are obviously higher than at lower redshift at present. In addition, they noticed that \textit{l}GRBs may be treated as a less biased probe of SFR at z>3 than at z<2 despite metallicity bias. Moreover, the discrepancy between GRB rate and the SFR at low redshift can be resulted from the physical effects of cosmic metallicity evolution \citep{2008MNRAS.388.1487L,2016ApJ...817....7P}, the low metallicity environment of host galaxies \citep{2014ApJS..213...15W,2016SSRv..202..181C}, etc. Notably, the majority of these low-redshift \textit{l}GRBs are less luminous as displayed in Figure~\ref{fig1}, which further indicates part of them could belong to a separate component with a Gauss-like profile in evidence.

\section{Conclusions}
\label{sec:summary}
We have carefully studied the effects of sample selection and threshold on the luminosity functions and event rates of distinct bright \textit{l}GRB samples and compared the GRB event rates with the SFRs in a more robust way. The following conclusions can be drawn:
\begin{itemize}
\item [1)]
The observed luminosity of \textit{l}GRBs in our samples evolves with the cosmological redshift as $L\propto(1+z)^k$ with an index $k$ varying from 2.88 to 3.92 that is marginally consistent with previous values. The $k$ parameter is more sensitive to the effects of sample selection instead of threshold of a detector if all bursts reside above the lower limits of luminosities.
\item [2)]
It is found for the first time that a Gaussian-like component of \textit{l}GRB event rates always exceeds the SFRs at lower redshift of $z<1$ no matter whether the selected GRB samples are complete or not. On the contrary, those high-redshift \textit{l}GRBs are perfectly associated with star formations, which is good in agreement with some previous conclusions. This directly demonstrates that two types of \textit{l}GRBs are evidently expected.
\item [3)]
It proves that the sample selection effect would play more important roles than the instrumental effect on calculating the cumulative luminosity functions, redshift distributions together with event rates of either complete or non-complete \textit{l}GRB samples. This is almost always right since the bursts located below the luminosity limits are very rare.
\item [4)]
It is worthy of addressing that the Gaussian-like excess of \textit{l}GRB event rates at $z<1$ in all three cases is largely different from the monotonous rise or drop patterns found before, which implies that the low-redshift \textit{l}GRBs might originate from some special progenitors unconnected with the SFRs at all. On the other hand, the high-redshift \textit{l}GRBs matching the SFRs ideally provide a convincing evidence supporting their physical origins from the core-collapse of massive stars.
\end{itemize}

\section{acknowledgements}

We are very grateful to the reviewer's comments that have greatly contributed to improving the quality of the publication. We acknowledge Y. F. Huang for constructive suggestion and discussions. We would like to thank F. Y. Wang, C. M. Deng and Y. Q. Qi for helpful comments. This work is partly supported by the National Natural Science Foundation of China (grant no U2031118), the Provincial Research Foundations (grant nos ZR2018MA030, XKJJC201901, and 201909118) and the science research grants from the China Manned Space Project with NO. CMS-CSST-2021-B11.

\section*{Data Availability}
The data underlying this article are available in the articles (Yu et al. 2015, ApJS, 218, 13; Zhang et al. 2018, PASP,130, 054202), Gamma-ray Coordinates Network (GCN) at https://gcn.gsfc.nasa.gov/ and The official Swift website https://swift.gsfc.nasa.gov/.

\clearpage
\onecolumn

\begin{longtable}{lcccccc}

    \caption{Typical parameters of 118 bright long GRBs}
	\label{tableGRB} \\
	\hline
	\hline
	\multicolumn{1}{c}{GRB}&\multicolumn{1}{c}{$z$}&\multicolumn{1}{c}{$T_{90}$}&\multicolumn{1}{c}{$P$}&\multicolumn{1}{c}{Energy band}&\multicolumn{1}{c}{$L$}&\multicolumn{1}{c}{Ref.}\\
	\multicolumn{1}{c}{}&\multicolumn{1}{c}{}&\multicolumn{1}{c}{$(s)$}&\multicolumn{1}{c}{$(\rm ph\ cm^{-2}~s^{-1})$}&\multicolumn{1}{c}{$(keV)$}&\multicolumn{1}{c}{$(\rm erg\ s^{-1})$}&\multicolumn{1}{c}{}\\
	\hline
	\endhead
050318$^*$ & 1.44  & 32    & 3.16  $\pm$0.2 & 15-150 & 4.96$\times10^{ 51}$ & 1,2,5,1,1 \\
    050401$^*$ & 2.9   & 33.3  & 10.7  $\pm$0.92 & 20-2000 & 2.09$ \times 10^{ 53}$ & 1,2,5,1,1 \\
    050416A$^*$ & 0.6535 & 2.5   & 4.88  $\pm$0.48 & 15-150 & 9.89$ \times 10^{ 50}$ & 1,2,5,1,1 \\
    050525A & 0.606 & 8.8   & 41.7  $\pm$0.94 & 15-150 & 9.00$ \times 10^{ 51}$ & 1,2,5,1,1 \\
    050603 & 2.821 & 12    & 21.5  $\pm$1.07 & 20-3000 & 2.25$ \times 10^{ 54}$ & 1,3,5,1,1 \\
    050802$^*$ & 1.71  & 19    & 2.75  $\pm$0.44 & 15-150 & 9.34$\times10^{51}$ & 1,2,5,1,1 \\
    050922C & 2.198 & 5     & 7.26  $\pm$0.32 & 20-2000 & 1.95$\times10^{53}$& 1,3,5,1,1 \\
    051109A & 2.346 & 37.2  & 3.94  $\pm$0.69 & 20-500 & 3.40$\times10^{52}$ & 1,2,5,1,1 \\
    051111 & 1.55  & 46    & 2.66  $\pm$0.21 & 15-350 & 7.04$\times10^{51}$ & 1,3,5,1,1 \\
    060206$^*$ & 4.048 & 7.6   & 2.79  $\pm$0.17 & 15-150 & 5.29$\times10^{52}$ & 1,2,5,1,1 \\
    060210$^*$ & 3.91  & 255   & 2.72  $\pm$0.28 & 15-150 & 5.64$\times10^{52}$ & 1,2,5,1,1 \\
    060306$^*$ & 3.5   & 60.94 & 5.97  $\pm$0.35 & 15-150 & 8.85$\times10^{52}$& 1,4,5,1,1 \\
    060614$^*$ & 0.125 & 108.7 & 11.5  $\pm$0.74 & 20-2000 & 2.33$\times10^{50}$ & 1,2,5,1,1 \\
    060814$^*$ & 0.84  & 145.3 & 7.27  $\pm$0.29 & 20-1000 & 9.46$\times10^{51}$ & 1,2,5,1,1 \\
    060908$^*$ & 1.8836 & 19    & 3.03  $\pm$0.25 & 15-150 & 1.34$\times10^{52}$ & 1,3,5,1,1 \\
    060927$^*$ & 5.47  & 22.5  & 2.7   $\pm$0.17 & 15-150 & 1.16$\times10^{53}$ & 1,2,5,1,1 \\
    061007$^*$ & 1.261 & 75.74 & 14.6  $\pm$0.37 & 20-10000 & 1.78$\times10^{53}$ & 1,4,5,1,1 \\
    061021$^*$ & 0.3463 & 47.82 & 6.11  $\pm$0.27 & 20-2000 & 1.76$\times10^{51}$ & 1,4,5,1,1 \\
    061121$^*$ & 1.314 & 81    & 21.1  $\pm$0.46 & 20-5000 & 1.48$\times10^{53}$  & 1,3,5,1,1 \\
    061222A$^*$ & 2.088 & 71.4  & 8.53  $\pm$0.26 & 20-10000 & 1.48$\times10^{53}$ & 1,2,5,1,1 \\
    070306$^*$ & 1.497 & 210   & 4.07  $\pm$0.21 & 15-150 & 1.04$\times10^{52}$ & 1,3,5,1,1 \\
    070508 & 0.82  & 20.9  & 24.1  $\pm$0.61 & 20-1000 & 2.96$\times10^{52}$ & 1,2,5,1,1 \\
    070714B & 0.92  & 3     & 2.7   $\pm$0.2 & 15-150 & 1.22$\times10^{52}$ & 1,3,5,1,1 \\
    071003 & 1.605 & 148   & 6.3   $\pm$0.4 & 20-4000 & 2.18$\times10^{53}$ & 1,3,5,1,1 \\
    071010B & 0.947 & 36.124 & 7.7   $\pm$0.3 & 20-1000 & 6.47$\times10^{51}$ & 1,2,5,1,1 \\
    071020$^*$ & 2.145 & 4.3   & 8.4   $\pm$0.3 & 20-2000 & 2.25$\times10^{53}$ & 1,4,5,1,1 \\
    071117$^*$ & 1.331 & 6.6   & 11.3  $\pm$0.4 & 20-1000 & 9.95$\times10^{52}$ & 1,2,5,1,1 \\
    080319B$^*$ & 0.937 & 125   & 24.8  $\pm$0.5 & 20-7000 & 1.05$\times10^{53}$ & 1,3,5,1,1 \\
    080319C & 1.95  & 34    & 5.2   $\pm$0.3 & 20-4000 & 9.46$\times10^{52}$ & 1,2,5,1,1 \\
    080411 & 1.03  & 56.33 & 43.2  $\pm$0.9 & 20-2000 & 9.33$\times10^{52}$ & 1,4,5,1,1 \\
    080413A & 2.433 & 46    & 5.6   $\pm$0.2 & 15-150 & 4.41$\times10^{52}$ & 1,2,5,1,1 \\
    080603B$^*$ & 2.69  & 60    & 3.5   $\pm$0.2 & 20-1000 & 1.11$\times10^{53}$ & 1,2,5,1,1 \\
    080605$^*$ & 1.6398 & 20    & 19.9  $\pm$0.6 & 20-2000 & 3.33$\times10^{53}$& 1,2,5,1,1 \\
    080607$^*$ & 3.036 & 78.97 & 23.1  $\pm$1.1 & 20-4000 & 2.21$\times10^{54}$  & 1,4,5,1,1 \\
    080721$^*$ & 2.602 & 176   & 20.9  $\pm$1.8 & 20-7000 & 1.11$\times10^{54}$  & 1,3,5,1,1 \\
    080804$^*$ & 2.2   & 37.87 & 3.1   $\pm$0.4 & 8-35000 & 2.86$\times10^{52}$ & 1,4,5,1,1 \\
    080916A$^*$ & 0.689 & 46.337 & 2.7   $\pm$0.2 & 8-35000 & 1.08$\times10^{51}$ & 1,2,5,1,1 \\
    081121$^*$ & 2.512 & 41.985 & 5.16  $\pm$1.53 & 8-35000 & 1.38$\times10^{53}$ & 1,2,5,1,1 \\
    081203A$^*$ & 2.1   & 223   & 2.9   $\pm$0.2 & 15-350 & 2.63$\times10^{52}$ & 1,3,5,1,1 \\
    081222$^*$ & 2.77  & 18.88 & 7.7   $\pm$0.2 & 8-35000 & 1.01$\times10^{53}$ & 1,2,5,1,1 \\
    090102$^*$ & 1.547 & 27    & 5.5   $\pm$0.8 & 8-35000 & 4.79$\times10^{52}$ & 1,2,5,1,1 \\
    090424$^*$ & 0.544 & 14.144 & 71    $\pm$2   & 8-35000 & 1.14$\times10^{52}$ & 1,2,5,1,1 \\
    090516 & 4.109 & 123.074 & 5.3   $\pm$0.2 & 8-1000 & 8.70$\times10^{52}$ & 1,2,5,1,1 \\
    090618 & 0.54  & 113   & 38.9  $\pm$0.8 & 8-35000 & 1.87$\times10^{52}$ & 1,3,5,1,1 \\
    090715B$^*$ & 3     & 265   & 3.8   $\pm$0.2 & 20-2000 & 8.78$\times10^{52}$ & 1,3,5,1,1 \\
    090812$^*$ & 2.452 & 75    & 2.77  $\pm$0.28 & 100-1000 & 1.02$\times10^{53}$ & 1,3,5,1,1 \\
    090926B$^*$ & 1.24  & 55.553 & 3.2   $\pm$0.3 & 8-35000 & 4.46$\times10^{51}$ & 1,2,5,1,1 \\
    091018$^*$ & 0.971 & 4.4   & 10.3  $\pm$0.4 & 20-1000 & 4.90$\times10^{51}$ & 1,2,5,1,1 \\
    091020$^*$ & 1.71  & 24.256 & 4.2   $\pm$0.3 & 8-35000 & 3.44$\times10^{52}$ & 1,2,5,1,1 \\
    091127$^*$ & 0.49  & 8.701 & 46.5  $\pm$2.7 & 8-35000 & 7.71$\times10^{51}$ & 1,2,5,1,1 \\
    091208B$^*$ & 1.063 & 12.48 & 15.2  $\pm$1   & 8-35000 & 1.81$\times10^{52}$ & 1,2,5,1,1 \\
    100615A$^*$ & 1.398 & 37.377 & 8.3   $\pm$0.2 & 8-1000 & 1.06$\times10^{52}$ & 1,2,5,1,1 \\
    100621A$^*$ & 0.542 & 63.6  & 12.8  $\pm$0.3 & 20-2000 & 3.24$\times10^{51}$ & 1,2,5,1,1 \\
    100728A & 1.567 & 165.378 & 5.1   $\pm$0.2 & 20-10000 & 6.45$\times10^{52}$ & 1,2,5,1,1 \\
    100728B$^*$ & 2.106 & 10.24 & 3.5   $\pm$0.5 & 8-35000 & 1.97$\times10^{52}$ & 1,2,5,1,1 \\
    100816A & 0.8049 & 2.045 & 15.59 $\pm$0.25 & 10-1000 & 7.38$\times10^{51}$ & 1,2,5,1,1 \\
    100906A & 1.727 & 110.594 & 10.1  $\pm$0.4 & 20-2000 & 6.90$\times10^{52}$ & 1,2,5,1,1 \\
    101213A & 0.414 & 131.12 & 4.67  $\pm$0.32 & 10-1000 & 6.32$\times10^{50}$& 1,2,5,1,1 \\
    110205A$^*$ & 2.22  & 257   & 3.6   $\pm$0.2 & 20-1200 & 2.65$\times10^{52}$ & 1,2,5,1,1 \\
    110213A & 1.46  & 34.305 & 17.7  $\pm$0.5 & 10-1000 & 2.23$\times10^{52}$ & 1,2,5,1,1 \\
    110422A & 1.77  & 25.9  & 30.7  $\pm$1   & 20-2000 & 2.90$\times10^{51}$ & 1,2,5,1,1 \\
    110715A & 0.82  & 13    & 53.9  $\pm$1.1 & 20-10000 & 4.31$\times10^{52}$ & 1,2,5,1,1 \\
    110731A & 2.83  & 7.485 & 20.9  $\pm$0.5 & 10-1000 & 3.06$\times10^{53}$ & 1,2,5,1,1 \\
    110818A & 3.36  & 67.073 & 5     $\pm$1.4 & 10-1000 & 7.23$\times10^{52}$ & 1,2,5,1,1 \\
    111008A$^*$ & 4.9898 & 63.46 & 6.4   $\pm$0.7 & 20-2000 & 4.95$\times10^{53}$ & 1,2,5,1,1 \\
    111228A$^*$ & 0.714 & 99.842 & 27    $\pm$1   & 10-1000 & 6.67$\times10^{51}$ & 1,2,5,1,1 \\
    120119A$^*$ & 1.728 & 55.297 & 16.86 $\pm$0.39 & 10-1000 & 5.98$\times10^{52}$ & 1,2,5,1,1 \\
    120326A$^*$ & 1.798 & 11.776 & 3.1   $\pm$0.05 & 10-1000 & 5.91$\times10^{51}$ & 1,2,5,1,1 \\
    120327A & 2.813 & 63.53 & 3.9   $\pm$0.2 & 15-350 & 3.42$\times10^{52}$ & 1,4,5,1,1 \\
    120712A & 4.1745 & 22.528 & 3.5   $\pm$0.2 & 10-1000 & 1.35$\times10^{53}$ & 1,2,5,1,1 \\
    120802A$^*$ & 3.796 & 50    & 3     $\pm$0.2 & 15-350 & 3.51$\times10^{52}$ & 1,2,5,1,1 \\
    120811C & 2.671 & 24.34 & 4.1   $\pm$0.2 & 15-350 & 2.23$\times10^{52}$ & 1,4,5,1,1 \\
    120907A$^*$ & 0.97  & 5.76  & 4.3   $\pm$0.4 & 10-1000 & 2.53$\times10^{51}$ & 1,2,5,1,1 \\
    120909A & 3.93  & 115   & 3     $\pm$0.2 & 10-1000 & 7.65$\times10^{52}$ & 1,2,5,1,1 \\
    120922A & 3.1   & 182.275 & 3.4   $\pm$0.3 & 10-1000 & 3.02$\times10^{52}$ & 1,2,5,1,1 \\
    121128A & 2.2   & 17.344 & 17.9  $\pm$0.5 & 10-1000 & 6.67$\times10^{52}$ & 1,2,5,1,1 \\
    130215A & 0.597 & 65.7  & 3.5   $\pm$0.3 & 10-1000 & 2.16$\times10^{51}$ & 1,2,5,1,1 \\
    130408A & 3.758 & 28    & 4.9   $\pm$1   & 20-10000 & 6.12$\times10^{53}$ & 1,2,5,1,1 \\
    130420A$^*$ & 1.297 & 104.96 & 5.2   $\pm$0.4 & 10-1000 & 3.65$\times10^{51}$ & 1,2,5,1,1 \\
    130427A$^*$ & 0.3399 & 138.242 & 1052  $\pm$2   & 8-1000 & 1.90$\times10^{53}$ & 1,2,5,1,1 \\
    130505A$^*$ & 2.27  & 89.34 & 30    $\pm$3.1 & 20-1200 & 3.98$\times10^{54}$ & 1,4,5,1,1 \\
    130514A & 3.6   & 214.19 & 2.8   $\pm$0.3 & 15-350 & 3.82$\times10^{52}$ & 1,4,5,1,1 \\
    130610A & 2.092 & 21.76 & 4.5   $\pm$0.9 & 10-1000 & 2.55$\times10^{52}$ & 1,2,5,1,1 \\
    130612A & 2.006 & 7.424 & 4.1   $\pm$0.2 & 10-1000 & 9.48$\times10^{51}$ & 1,2,5,1,1 \\
    130701A$^*$ & 1.155 & 4.38  & 17.1  $\pm$0.7 & 20-10000 & 4.27$\times10^{52}$ & 1,2,5,1,1 \\
    130831A$^*$ & 0.4791 & 30.19 & 13.6  $\pm$0.6 & 20-10000 & 3.42$\times10^{51}$ & 1,4,5,1,1 \\
    130907A$^*$ & 1.238 & 364.37 & 25.6  $\pm$0.5 & 20-10000 & 1.82$\times10^{53}$ & 1,4,5,1,1 \\
    131030A$^*$ & 1.295 & 39.42 & 28.1  $\pm$0.7 & 20-10000 & 1.08$\times10^{53}$ & 1,4,5,1,1 \\
    \hline
    \hline
    070521$^*$ & 2.087 & 37.9  & 6.5   $\pm$0.26 & 15-150 & 1.44$\times10^{53}$ & 2,2,5,2,2 \\
    080430$^*$ & 0.767 & 16.2  & 2.65  $\pm$0.17 & 15-150 & 1.51$\times10^{51}$ & 2,2,5,2,2 \\
    081007$^*$ & 0.5295 & 10    & 2.9   $\pm$0.4 & 15-150 & 3.74$\times10^{50}$ & 2,2,5,2,2 \\
    090328$^*$ & 0.736 & 61.697 & 25.35 $\pm$1.5 & 10-1000 & 2.00$\times10^{52}$ & 2,2,5,2,2 \\
    091003$^*$ & 0.8969 & 20.224 & 46.63 $\pm$2.21 & 10-1000 & 4.73$\times10^{52}$ & 2,2,5,2,2 \\
    091024 & 1.09  & 93.954 & 5.65  $\pm$1.17 & 10-1000 & 1.34$\times10^{52}$ & 2,2,5,2,2 \\
    101219B$^*$ & 0.552 & 51.009 & 3.16  $\pm$0.84 & 10-1000 & 5.58$\times10^{50}$ & 2,2,5,2,2 \\
    120729A$^*$ & 0.8   & 25.472 & 6.61  $\pm$1.28 & 10-1000 & 5.92$\times10^{51}$ & 2,2,5,2,2 \\
    121211A$^*$ & 1.023 & 5.632 & 4.3   $\pm$1.07 & 10-1000 & 2.30$\times10^{51}$ & 2,2,5,2,2 \\
    130702A$^*$ & 0.145 & 58.881 & 16.51 $\pm$4.69 & 10-1000 & 5.75$\times10^{49}$ & 2,2,5,2,2 \\
    130925A$^*$ & 0.35  & 6.4   & 5.88  $\pm$1.66 & 10-1000 & 1.58$\times10^{50}$ & 2,2,5,2,2 \\
    131231A$^*$ & 0.642 & 31.232 & 82.72 $\pm$2.68 & 10-1000 & 2.27$\times10^{52}$ & 2,2,5,2,2 \\
    140506A$^*$ & 0.889 & 64.128 & 24.8  $\pm$2.94 & 10-1000 & 2.02$\times10^{52}$ & 2,2,5,2,2 \\
    140508A$^*$ & 1.027 & 44.288 & 88.15 $\pm$4.32 & 10-1000 & 8.97$\times10^{52}$ & 2,2,5,2,2 \\
    140606B$^*$ & 0.384 & 22.784 & 16.26 $\pm$1.35 & 10-1000 & 2.30$\times10^{51}$ & 2,2,5,2,2 \\
    141004A & 0.57  & 9.472 & 12.02 $\pm$1.2 & 10-1000 & 1.47$\times10^{51}$ & 2,2,5,2,2 \\
    141225A$^*$ & 0.915 & 56.32 & 4.42  $\pm$1.13 & 10-1000 & 5.46$\times10^{51}$ & 2,2,5,2,2 \\
    150323A$^*$ & 0.593 & 149.6 & 4.56  $\pm$0.24 & 15-150 & 1.09$\times10^{51}$ & 2,2,5,2,2 \\
    150514A$^*$ & 0.807 & 10.813 & 22.93 $\pm$1.56 & 10-1000 & 6.92$\times10^{51}$ & 2,2,5,2,2 \\
    150727A$^*$ & 0.313 & 49.409 & 4.67  $\pm$1.22 & 10-1000 & 4.19$\times10^{50}$ & 2,2,5,2,2 \\
    150821A$^*$ & 0.755 & 103.426 & 13.56 $\pm$1.41 & 10-1000 & 7.32$\times10^{51}$ & 2,2,5,2,2 \\
    151001B & 0.093 & 23.552 & 3.71  $\pm$1.26 & 10-1000 & 2.67$\times10^{49}$ & 2,2,5,2,2 \\
    151027A$^*$ & 0.81  & 123.394 & 13.08 $\pm$1.37 & 10-1000 & 8.54$\times10^{51}$ & 2,2,5,2,2 \\
    160425A$^*$ & 0.555 & 304.58 & 2.8   $\pm$0.2 & 15-150 & 8.59$\times10^{50}$ & 2,2,5,2,2 \\
    160623A & 0.367 & 107.776 & 4.26  $\pm$1.36 & 10-1000 & 7.24$\times10^{50}$ & 2,2,5,2,2 \\
    160804A$^*$ & 0.736 & 131.586 & 7.24  $\pm$1.64 & 10-1000 & 1.54$\times10^{51}$ & 2,2,5,2,2 \\
    161129A & 0.645 & 36.096 & 11.92 $\pm$1.45 & 10-1000 & 4.54$\times10^{51}$ & 2,2,5,2,2 \\
    161219B$^*$ & 0.1475 & 6.94  & 4.4   $\pm$0.4 & 15-150 & 2.67$\times10^{49}$ & 2,2,5,2,2 \\
    170903A$^*$ & 0.886 & 25.6  & 7.3   $\pm$1.59 & 15-150 & 4.28$\times10^{51}$ & 2,2,5,2,2 \\
    171010A$^*$ & 0.3285 & 107.266 & 137.3 $\pm$4.41 & 10-1000 & 7.99$\times10^{51}$ & 2,2,5,2,2 \\
	\hline	
\end{longtable}
	\footnotesize
	{Note. The first set lists the 88 long GRBs in sample I and the second set represents 30 low-redshift long bursts taken from \cite{2018PASP..130e4202Z}. All 118 GRBs are used to build the sample II of this work. The references are given in order for redshift, duration, peak photon flux, energy range of a detector and bolometric luminosity of a given GRB.
1.~\citet{2015ApJS..218...13Y};2.~\citet{2018PASP..130e4202Z}; 3.~\citet{2012ApJ...746..156C};4.~\citet{2020ApJS..248...21H};5.~https://www.mpe.mpg.de/~jcg/grbgen.html\\
$^*$These GRBs satisfying the selection conditions \citep{2006A&A...447..897J,2012ApJ...749...68S} constitute our complete sample III.}

\bsp	
\label{lastpage}

\bibliographystyle{mnras}

\end{document}